\documentclass[apj]{emulateapj}
\usepackage{apjfonts}
\usepackage{epsfig}
\usepackage{graphicx}
\usepackage{url}
\usepackage{natbib}
\usepackage{float}
\usepackage{amsmath}
\usepackage[bf, sf, FIGTOPCAP, nooneline, tight]{subfigure}
\usepackage{graphicx}
\usepackage{dcolumn}
\usepackage{bm}
\usepackage[usenames,dvipsnames]{color}
\usepackage[citebordercolor={0 1 0}]{hyperref}
\usepackage{amssymb}
\usepackage{ulem}

\newcommand{\be}{\begin{equation}}
\newcommand{\ee}{\end{equation}}
\newcommand{\bq}{\begin{eqnarray}}
\newcommand{\eq}{\end{eqnarray}}

\def\({\left(}
\def\){\right)}

\begin{document}

\title{Cosmic Velocity Field Reconstruction Using AI}

\author{
  Ziyong Wu\altaffilmark{1},
  Zhenyu Zhang \altaffilmark{1},
  Shuyang Pan\altaffilmark{1},
  Haitao Miao\altaffilmark{1}, Xiaolin Luo\altaffilmark{1},
  Xin Wang\altaffilmark{1 $^\dagger$},
  Cristiano G. Sabiu\altaffilmark{2,3},
  Jaime Forero-Romero\altaffilmark{4},
  Yang Wang\altaffilmark{1 $^\ddagger$},
  Xiao-Dong Li\altaffilmark{1 $^\star$}
}

\email{$\dagger$wangxin35@mail.sysu.edu.cn}
\email{$^\ddagger$wangyang23@mail.sysu.edu.cn}
\email{$^\star$lixiaod25@mail.sysu.edu.cn}


\affil{School of Physics and Astronomy, Sun Yat-Sen University, Guangzhou 510297, P. R. China$^1$}
\affil{Department  of  Astronomy,  Yonsei  University,  50  Yonsei-ro,Seoul 03722, Korea$^2$} {}
\affil{Natural Science Research Institute, University of Seoul, 163 Seoulsiripdaero, Dongdaemun-gu, Seoul, 02504, Republic of Korea$^3$} {}
\affil{Departamento de F{\'i}sica, Universidad de los Andes, Cra. 1 No. 18A-10 Edificio Ip, CP 111711, Bogot{\'a}, Colombia$^4$}


\begin{abstract}
  We develop a deep learning technique to infer the
  non-linear velocity field from the dark matter density field.
  The deep learning architecture we use is an  ``U-net'' style convolutional neural network,
  which consists of 15 convolution layers and 2 deconvolution layers.
  This setup maps the 3-dimensional density field of $32^3$-voxels to the 3-dimensional velocity or momentum fields of $20^3$-voxels.
  Through the analysis of the dark matter simulation with a resolution of $2 {h^{-1}}{\rm Mpc}$,
  we find that the network can predict the the non-linearity, complexity  and vorticity of the velocity and momentum fields,
  as well as the power spectra of their value, divergence and vorticity and its prediction accuracy reaches the range of $k\simeq1.4$ $h{\rm Mpc}^{-1}$
  with a relative error ranging from 1\% to $\lesssim$10\%.
  A simple comparison shows that neural networks may have an
  overwhelming advantage over perturbation theory in the
  reconstruction of velocity or momentum fields.
  %
\end{abstract}

\maketitle

\section{Introduction}\label{intro}
The large-scale structure (LSS) of the Universe is a key observational probe to
study the physics of dark matter, dark energy, gravity and cosmic neutrinos.
In the next 10 years, stage IV surveys, including DESI \footnote{https://desi.lbl.gov/},
EUCLID \footnote{http://sci.esa.int/euclid/}, LSST \footnote{http://sci.esa.int/euclid/},
WFIRST \footnote{https://wfirst.gsfc.nasa.gov/}, and CSST, will begin
to map out an unprecedented large volume of the Universe with
extraordinary precision.
It is of critical importance to have statistical tools that can
reliably extract the physical information in the LSS data.

The peculiar velocities of the galaxies, sourced by the ``initial''
inhomogeneities, is an excellent probe for the physics of the LSS,
enabling us to better study or measure such quantities as the redshift
space distortions \cite{kaiser1987clustering,jackson1972critique},
baryon acoustic oscillations
\citep{Eisenstein:2005su,Eisenstein2007BAOReconstruction}, the
Alcock-Paczynski effect \citep{ap,Li2014,Li2015,Li2016,KR2018}, the
cosmic web
\citep{1986Bardeen,hahn2007properties,forero2009dynamical,hoffman2012kinematic,forero2014cosmic,Fang2019},
the kinematic Sunyaev-Zeldovich effect \citep{1972CoASP...4..173S,
  1980MNRAS.190..413S}, and the integrated Sachs Wolfe effect
\citep{1967ApJ...147...73S, 1968Natur.217..511R,
  1996PhRvL..76..575C}.

Observationally, the measurement  of the peculiar velocities is a difficult task,
as it requires redshift independent determination of the distance,
which is usually accomplished via distance indicators such as
type Ia Supernovae \citep{1993ApJ...413L.105P,SNIaflow...1997ApJ...488L...1R,
  SNIaflow...2004MNRAS.355.1378R,SNIaflow...2012MNRAS.420..447T,SNIaflow...2016ApJ...827...60M}
the Tully-Fisher relation \citep{TullyFisher...1977A&A....54..661T,
  TullyFisher...2006ApJ...653..861M,TullyFisher...2008AJ....135.1738M}
and the Fundamental Plane relation \citep{FundPlan...1987ApJ...313...42D,
  FundPlan...1987ApJ...313...59D,FundPlan...2007ApJS..172..599S}
As an alternative approach,
one can ``reconstruct'' the cosmic velocity field from the density field
based on their relationship described by theories.
Here the difficulty is the complexity caused by the non-linear evolution of the structures.
Numerous works have been done in this direction.
For more details, one can check
\cite{VelocityRecon...1991ApJ...379....6N,
  VelocityRecon...1992ApJ...390L..61B,
  VelocityRecon...1995ApJ...449..446Z,
  VelocityRecon...1997MNRAS.285..793C,
  VelocityRecon...1999MNRAS.309..543B,
  VelocityRecon...2000MNRAS.316..464K,
  VelocityRecon...2002MNRAS.335...53B,
  VelocityRecon...2005ApJ...635L.113M,
  VelocityRecon...2008MNRAS.383.1292L,
  VelocityRecon...2008MNRAS.391.1796B,
  VelocityRecon...2012MNRAS.425.2422K,
  VelocityRecon...2012MNRAS.420.1809W,
  VelocityRecon...2015MNRAS.449.3407J,2017MNRAS.467.3993A}.

Recently machine learning algorithms, especially those based on deep neural networks,
are becoming promising toolkits for the study of complex data
that are difficult to be solved by traditional methods.
So far, this technique have been applied to almost all sub-fields of cosmology, including weak gravitational lensing \citep{Schmelzle:2017vwd,
  Gupta:2018eev,Springer:2018aak,Fluri:2019qtp,Jeffrey:2019fag,Merten:2018bgr,Peel:2018aei,Tewes:2018she},
the cosmic microwave background \citep{Caldeira:2018ojb,Rodriguez:2018mjb,Perraudin:2018rbt,Munchmeyer:2019kng,Mishra:2019sep},
the large scale structure \citep{Ravanbakhsh:2017bbi,Lucie-Smith:2018smo,
  Modi:2018cfi,Berger:2018aey,He:2018ggn,Lucie-Smith:2019hdl,
  Pfeffer:2019pca,Ramanah:2019cbm,Troster:2019mys,Zhang:2019ryt,AIBAORecon...2020arXiv200210218M,Li2020...ML...2020SCPMA..63k0412P},
gravitational waves \citep{Dreissigacker:2019edy,Gebhard:2019ldz},
cosmic reionization \citep{LaPlante:2018pst,Gillet:2018fgb,Hassan:2018bbm,Chardin:2019euc,Hassan:2019cal},
supernovae \citep{Lochner:2016hbn,Moss:2018tug,Ishida:2018uqu,Li:2019ybe,Muthukrishna:2019wpf}.
For more details, one can refer to \cite{Mehta:2018dln,Jennings:2018eko,Carleo:2019ptp,Ntampaka:2019udw} and the references therein.


In this paper, we apply deep learning techniques to reconstruct the
velocity field from the dark matter density field.
This converts the reconstruction problem to a non-linear mapping between the two fields,
which is achieved via a deep neural network with a U-net style architecture.
This paper is organized as follows. In section 2, we introduce the dataset
and data processing methods we use. In section 3, we  discuss our
neural network,  including the construction of our neural network, the selection
of parameters and details of training, etc. Section 4 presents the main results, and
section 5 represents the conclusion and discussion.

\section{TRAINING AND TESTING DATASETS}\label{dataset}


The training and testing samples are generated using the COmoving Lagrangian Acceleration (COLA) code \citep{2013JCAP...06..036T}.
COLA computes the evolution of dark matter particles in a frame that is comoving with observers following trajectories predicted by the Lagrangian Perturbation Theory (LPT),
in order to accurately deal with the small-scale structures, without sacrificing the accuracy of large scales.
Being hundreds of times faster than N-body simulations, it still maintains a good accuracy from very large to highly non-linear scales.

We generate a set of 14 simulations, assuming a $\Lambda$CDM
cosmology $\Omega_{m}=0.31$, $\Omega_{b}=0.05$,
$\sigma_{8}=0.83$, $n_{s}=0.96$, $H_{0}=67.77$km$\cdot $s$^{-1}$
Mpc$^{-1}$.
Each of the simulation is run  within a cube with a volume of
$(512\ h^{-1}{\rm Mpc} )^{3}$ using $512^{3}$ dark matter
particles, 
having a mean separation of 1 $h^{-1}$Mpc  per dimension.
The output at $z=0$ are then used for the main part of our analysis.

\begin{figure*}[htbp]
  \centering
  \includegraphics[width=8cm,height=5cm]{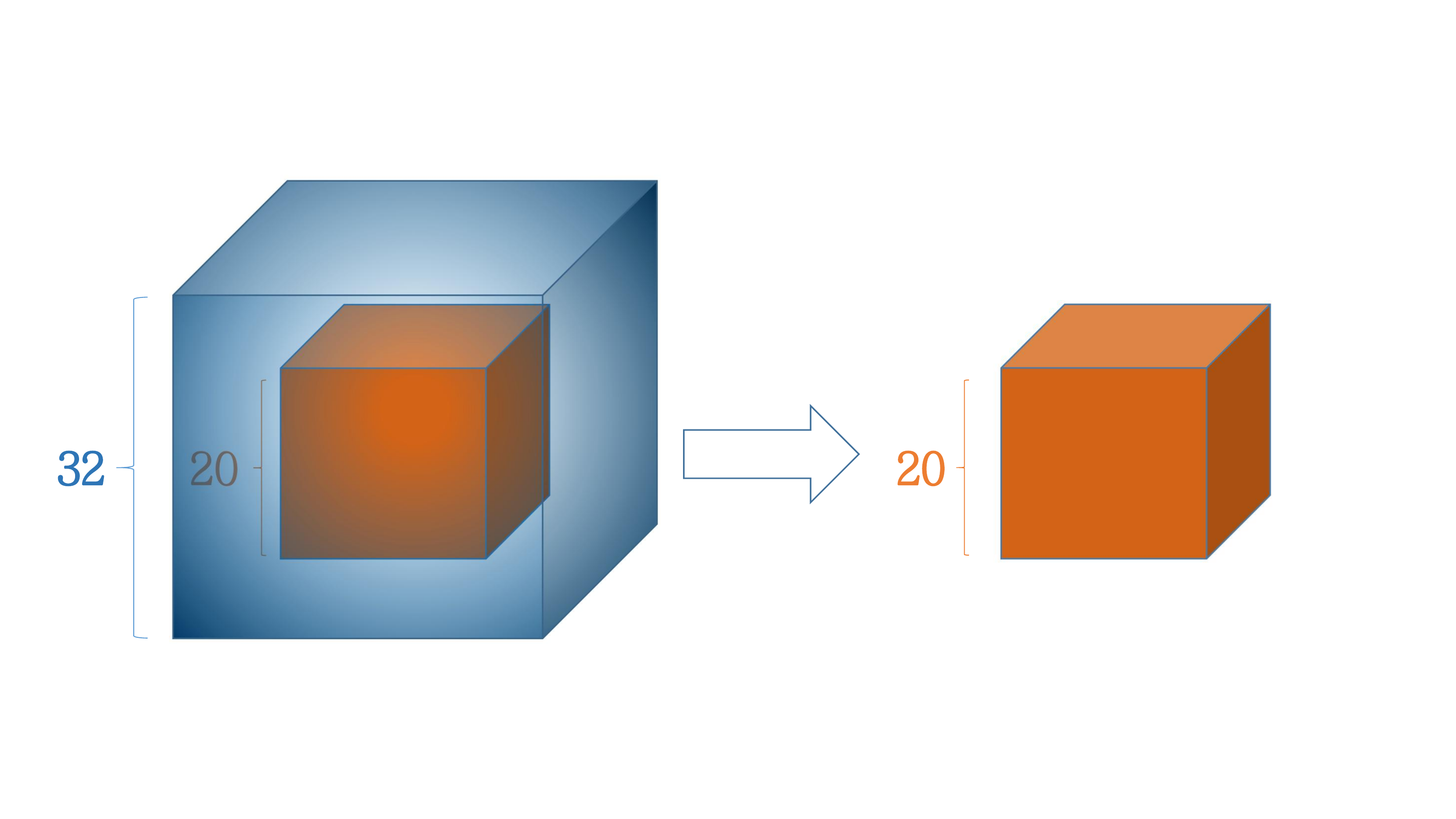}
  \includegraphics[width=8cm,height=5cm]{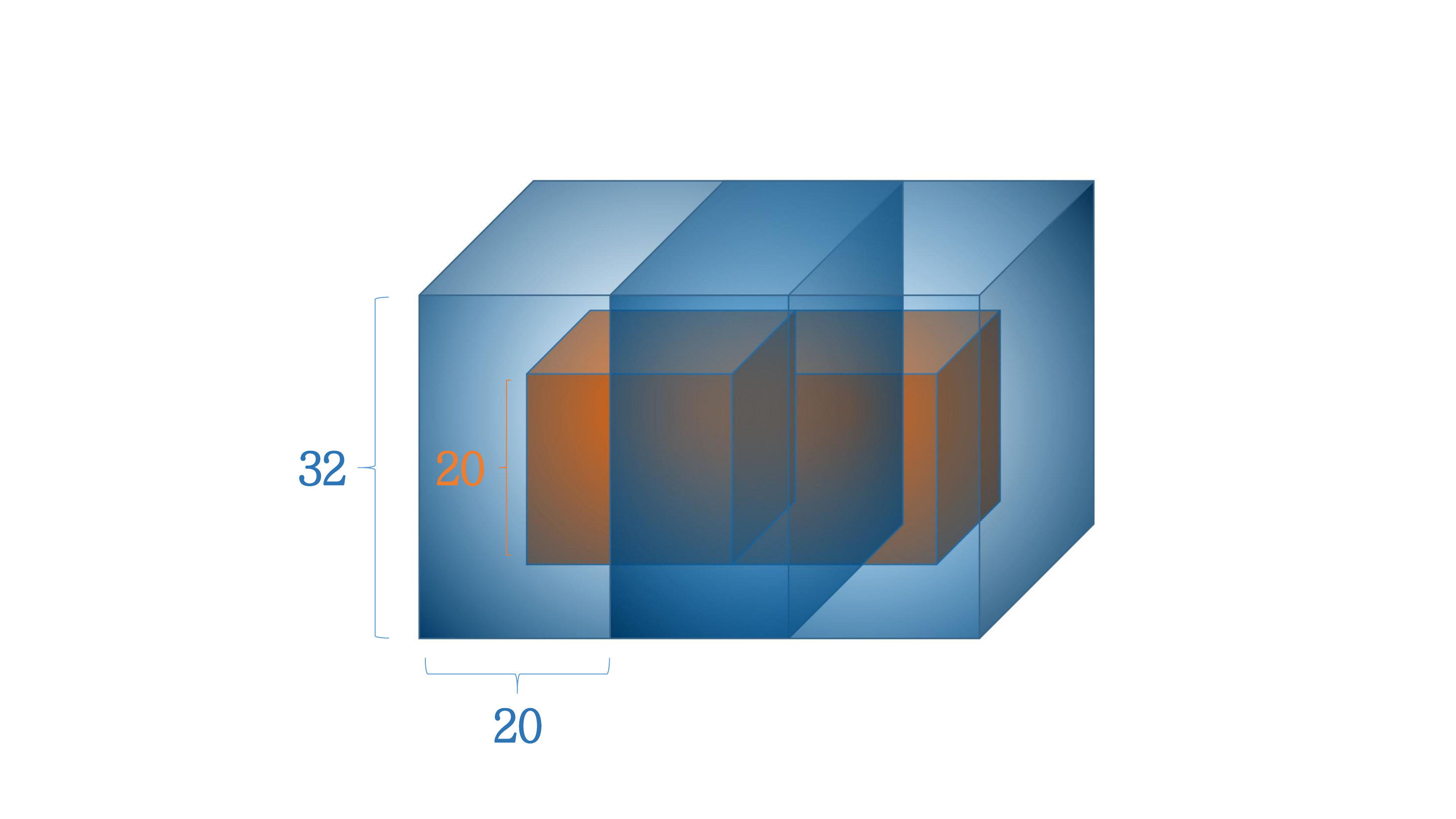}
  \caption{Left panel: The input of our neural network is a $32^3$-voxel density field (blue), while
    the output is a $20^3$-voxel velocity field (red) located around the center of the input.
    This choice reduces boundary effects.
    Right panel: A series of overlapped input fields yield to
    non-overlapped outputs which can be spliced back to build up
    larger cubes.}
  \label{fig:subcube}
\end{figure*}
\begin{figure*}[htbp]
  \centering

  \includegraphics[width=16cm,height=10cm]{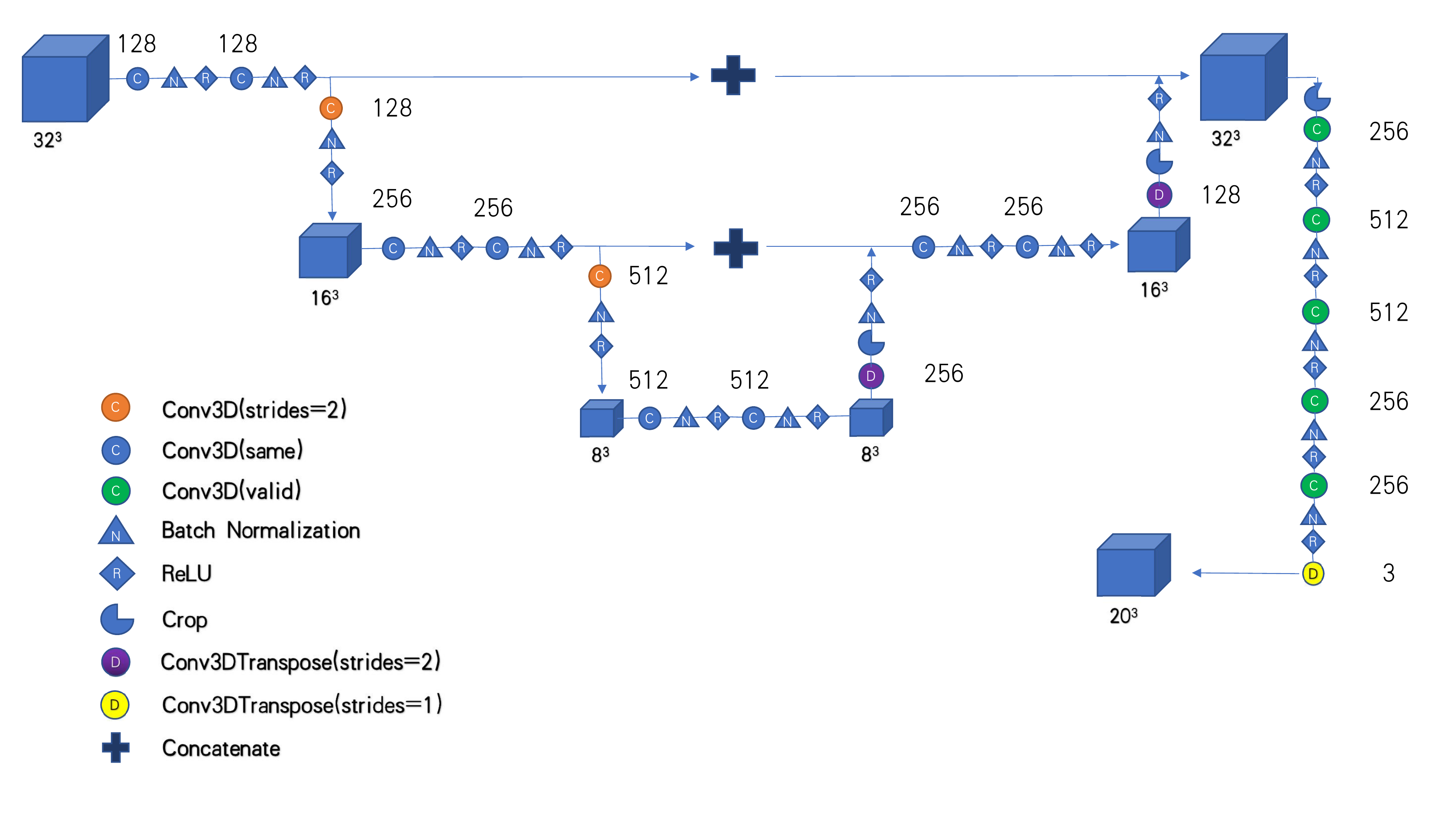}
  \caption{The overall structure of our network is designed similar to the ``U-net'' style architecture,
    which is built upon the Convolutional Network
    and modified in a way that it has better performance in imaging analysis.
    It is consisting of 15 convolution layers and 2 deconvolution layers,
    and maps the $32^{3}$-voxel density field($64\ h^{-1} {\rm Mpc}$) to the $20^{3}$-voxel velocity or momentum fields($40\ h^{-1} {\rm Mpc}$).
    A lot of detailed designs are adopted to guarantee the performance of the network.}
  \label{fig:unet}
\end{figure*}

The Clouding-In-Cells (CIC) algorithm is adopted for
constructing the density and momentum fields from the outputs.
Since the momentum has three dimensions, for each sample we need
to construct three fields describing $p_x$, $p_y$ and $p_z$,
respectively.
The division of the momentum and density fields then leads to
three velocity fields, i.e. $v_x(\bf x)$, $v_x(\bf x)$ and
$v_z(\bf x)$
\footnote{One small problem is that at some lattice points the value of the density is estimated to be zero.
  We assign them the background velocity, which equals to the mean momentum divided by the mean density.
}.
For all fields, we choose a resolution of ($2$ $h^{-1}$Mpc
)$^{3}$, corresponding to $256^3$ voxels.




In practice, we further split the density
and momentum/velocity voxels into smaller sub-cubes before
feeding them to the neural network.
We take such process based on the following considerations:
\begin{itemize}
  \item
        Learning a larger cube requires a larger number of neurons or layers in the network,
        making the training more difficult and expensive.
  \item
        Dealing with large fields is limited by memory constraints,
        especially if GPUs are used in the training process.
  \item
        By using small cubes as training samples,
        we force the neural network to focus on interpreting
        and predicting the small-scale, non-linear patterns in the velocity fields.
        The large-scale velocity field, which can be easily estimated using perturbation theory,
        is not our focus.
\end{itemize}

\begin{figure*}[htbp]
  \centering
  \includegraphics[width=16cm,height=7.5cm]{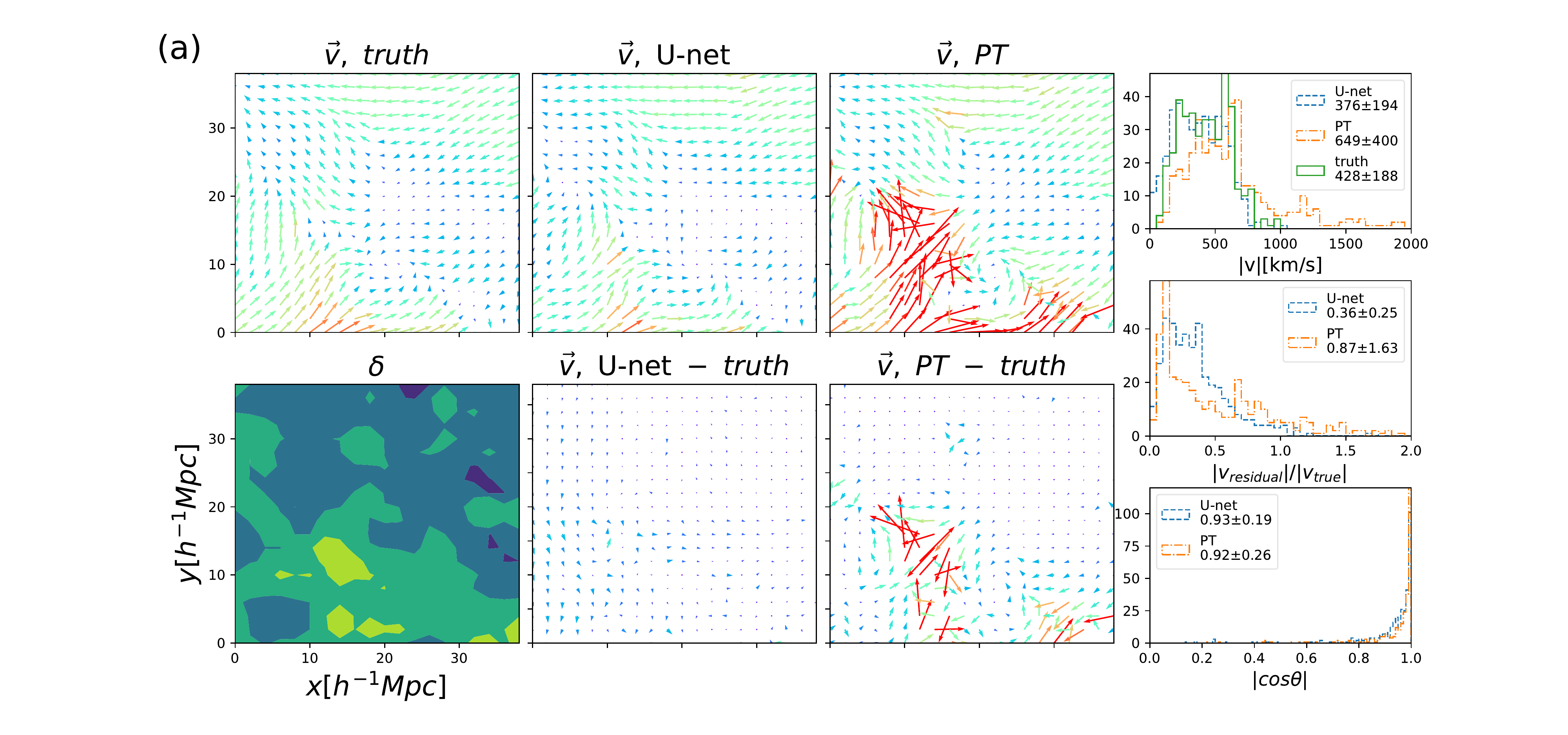}
  \includegraphics[width=16cm,height=7.5cm]{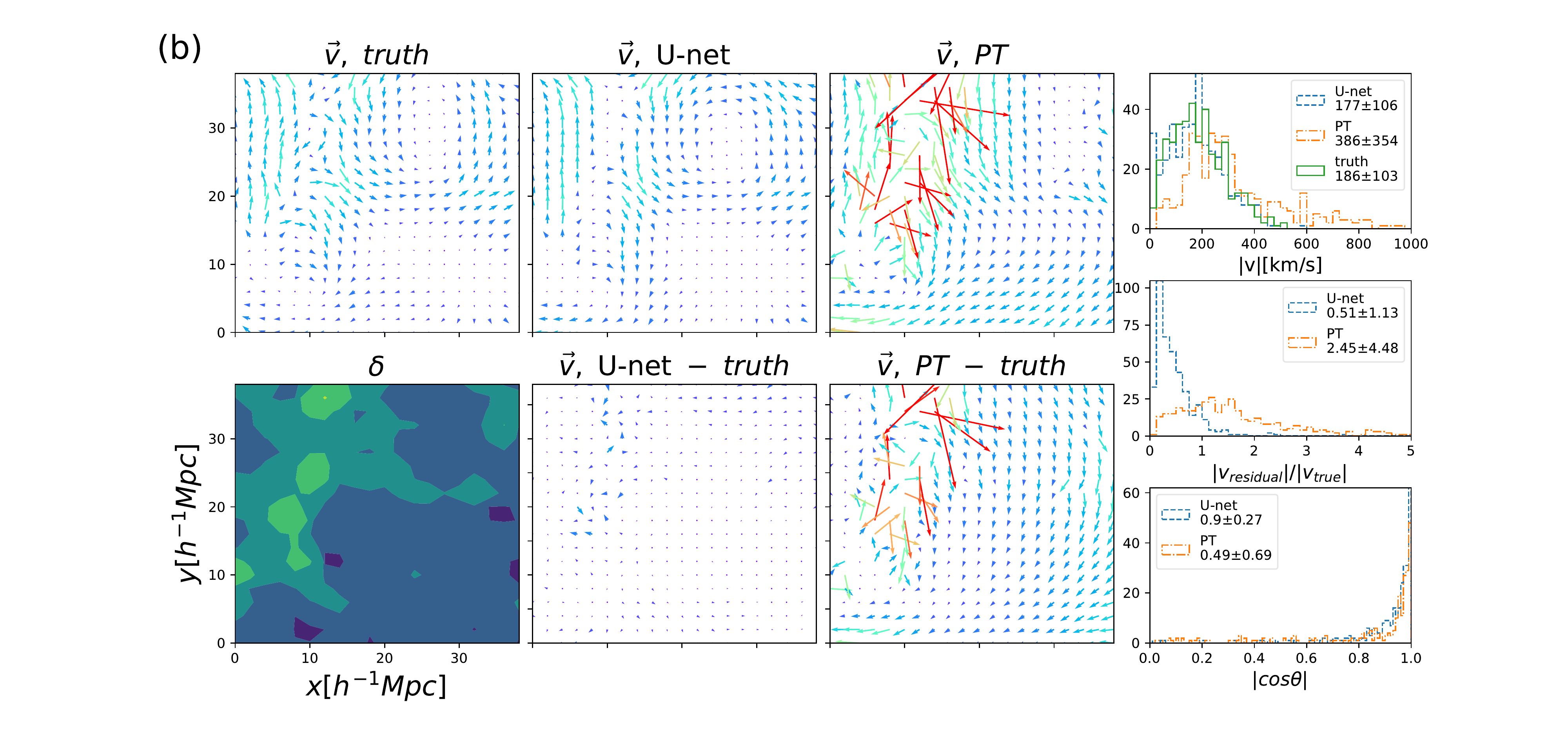}
  \includegraphics[width=16cm,height=7.5cm]{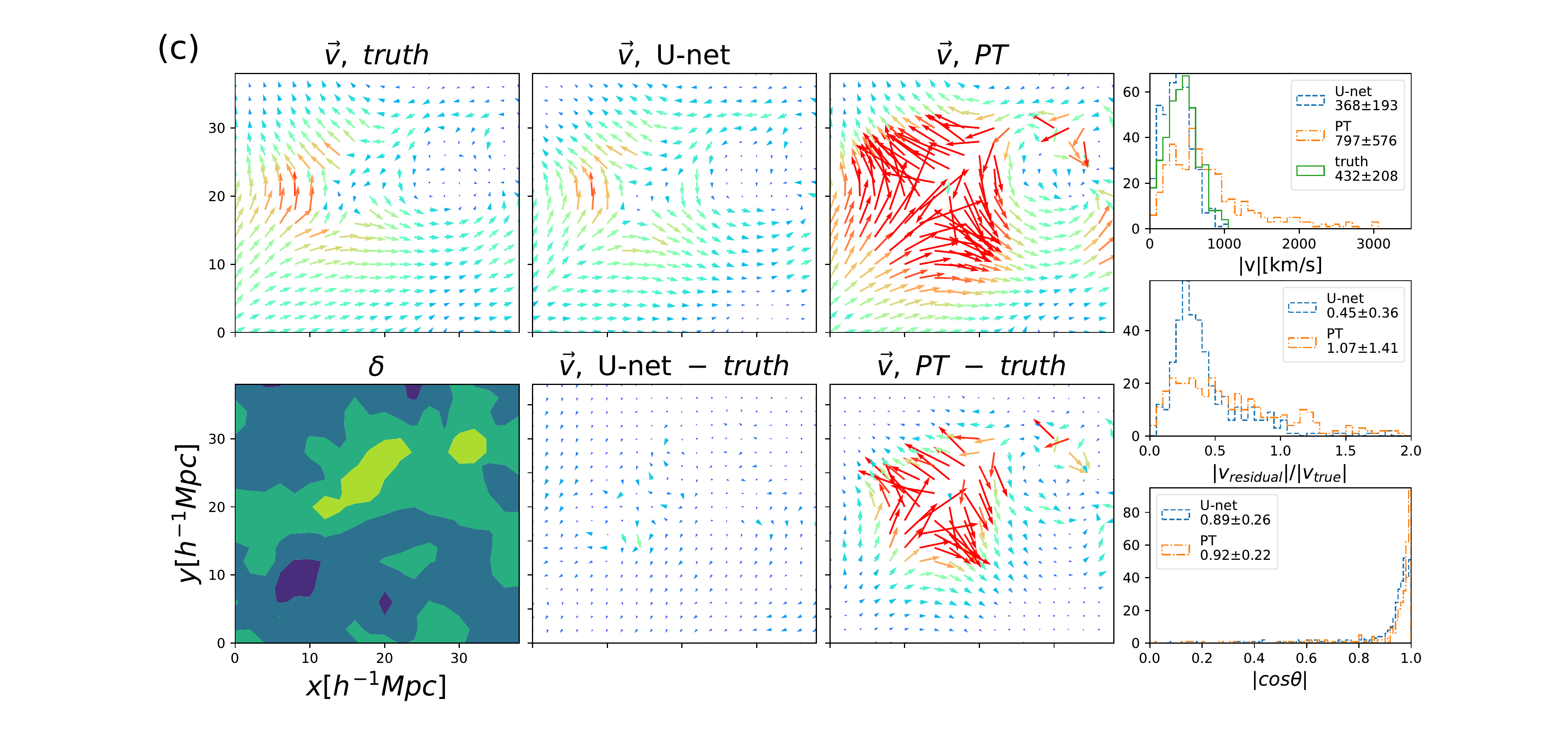}
  \caption{
  Three slices selected from the testing sample.
  In regions where two bulks of matter collide and merge,
  the velocity is highly non-linear.
  For each slice, We plot the velocity field of original input (top left),U-net prediction(top middle) and perturbation theory prediction(top right). In the bottom row, we also plot the corresponding density field(bottom left), residual of U-net prediction(bottom middle) and PT prediction(bottom right).
  In the right column, we plot the histograms of $v$, $|v_{\rm resiual}|/|v_{\rm true}|$ and $|\cos \theta|$. All these suggest that the performance of neural network is much better.
  \label{fig:velocities}
  }
\end{figure*}

\begin{figure*}[htbp]
  \centering
  \includegraphics[width=16cm,height=7.5cm]{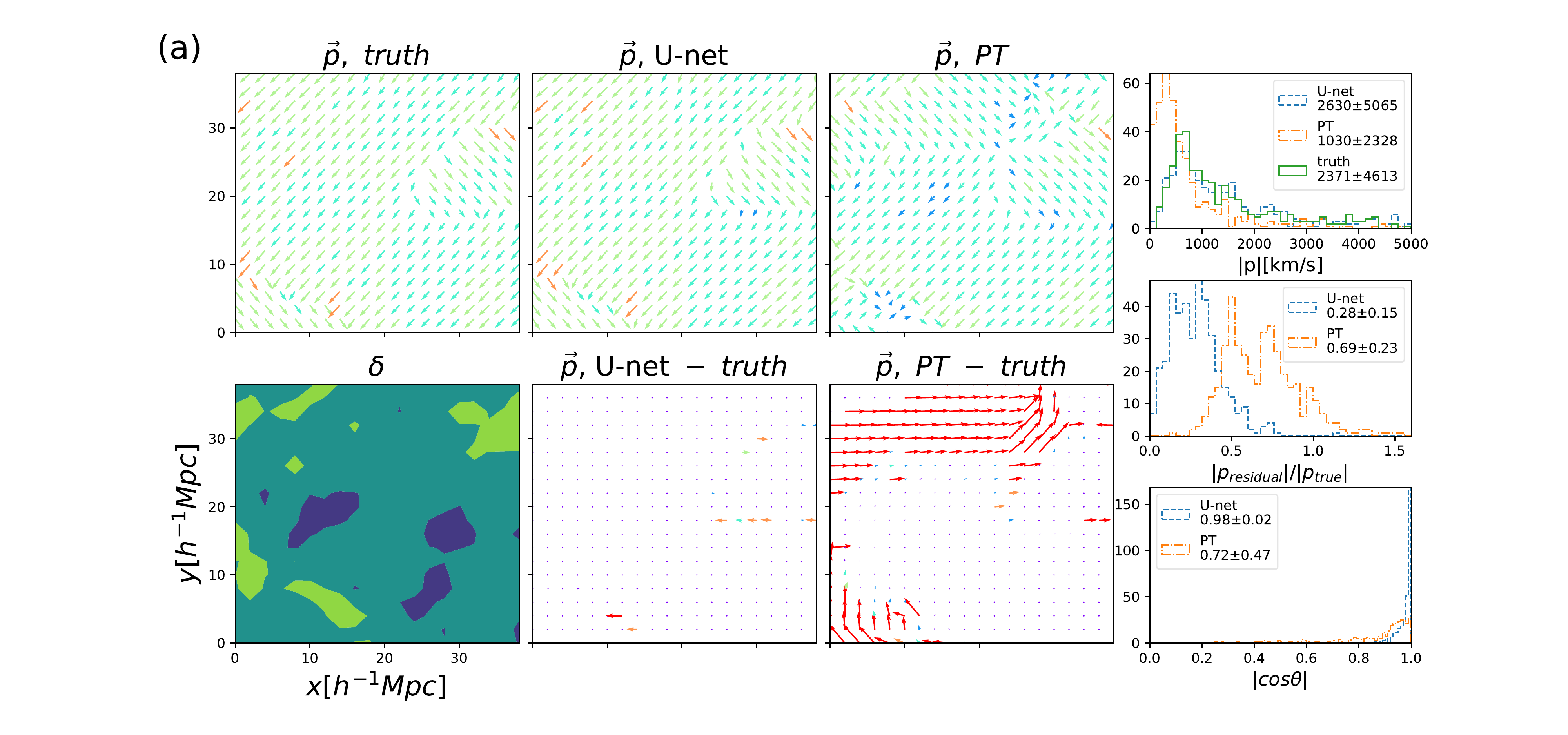}
  \includegraphics[width=16cm,height=7.5cm]{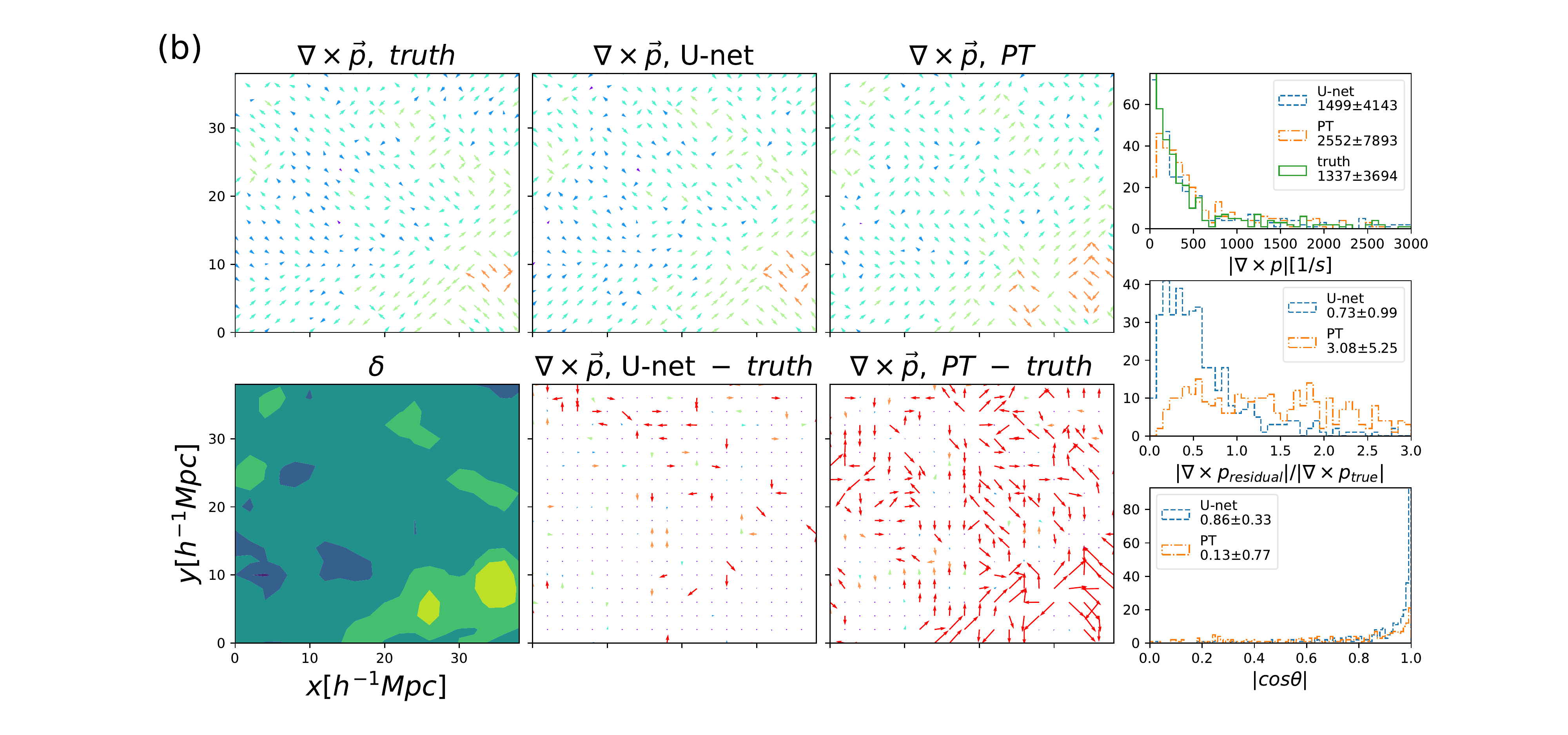}
  \includegraphics[width=16cm,height=7.5cm]{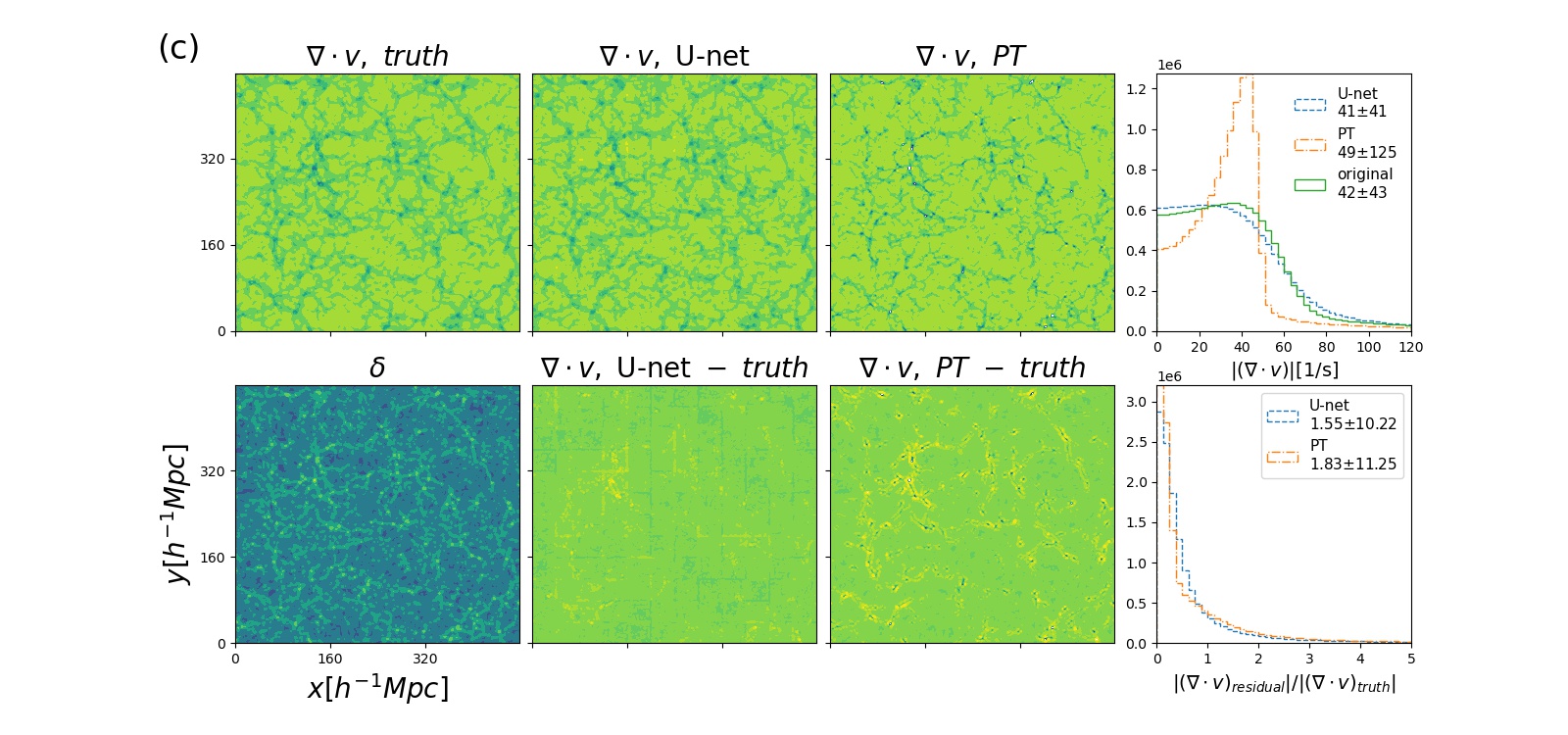}
  \caption{
    We compare the momentum, momentum curl and velocity divergence.Similar to Fig \ref{fig:velocities}, we plot the field of original input (top left),U-net prediction(top middle), perturbation theory prediction(top right),
    corresponding density field(bottom left), residual of U-net prediction(bottom middle) and PT prediction(bottom right) as well as the histograms of quantities, their residuals and angle.
    We find the U-net prediction is almost the same as the truth one,
    while the linear perturbation theory loss many detail structures. 
    These all suggest that the performance of neural network is much better.
  }
  \label{fig:momentum}
\end{figure*}



To avoid possible inaccuracy and complexity brought by the
boundary effects, the neural network is designed to map the
density fields into momentum fields {\it having a smaller
    size}.
For each momentum filed, we take a $240^{3}$-voxel subfield from it,
cut the subfield into 1,728 $20^{3}$-voxel subcubes,
and set the subcubes as the targets (i.e. outputs) of the neural network.
The inputs of the network are a series of $32^3$-voxel density
fields sharing the same centers with those momentum fields.
In this way, 75\% voxels (lying near the outer boundary of the density fields) serve as adjacent points,
for the purpose of enhancing accuracy\ref{fig:subcube}.

Furthermore, since the density values span three orders of magnitude,
it is difficult for the neural network to establish an accurate
mapping. Thus we use the following logarithmic transform to mitigate this
problem
\begin{equation}
  \tilde{\rho}(x,y,z)=\ln(\rho (x,y,z)+1).
\end{equation}

By using their log values, we greatly decrease the variance.
Moreover,  the distribution of large scale structure density is
close to the lognormal distribution,
so we can use the above expression to convert it into an
approximate normal distribution\cite{2012ApJ...745...17F,2009ApJ...698L..90N,2012MNRAS.425.2443K}.

The $32^3$-voxel fields are split into training, verification and test
sets,   among which the training set accounts for 60\%, the
verification set accounts for 30\%,  and the test set accounts for
10\% of the total data.
The single batch number for the training is set as 6. 

\section{NEURAL NETWORK ARCHITECTURE}\label{NETWORK}

We adopt a ``U-net'' style architecture,
which is built upon the Convolutional Network
and modified in a way that it has better performance in imaging
analysis.

As mentioned in the previous section, the entire simulation box were
into $32^{3}$-voxel subcubes with a size of $(64\ h^{-1} {\rm Mpc})^{3}$,
and mapped into $20^{3}$-voxel velocity fields.
$64$ $h^{-1}$Mpc  is large enough to capture the non-linear features in the field,
while reducing the input data complexity, thus reducing the required number of
neurons.
Accordingly, the overall structure of our network is designed as follows (see Figure \ref{fig:unet}),
\begin{itemize}
  \item First, the input $32$-voxel density fields are fed into two convolution layers,
        which {\it convolve} the inputs and pass the resulting feature fields to the next-level layers.
        To capture the abundant features in the 3-D LSS, each layer has 128 filters,
        while each filter has a shape of $3^3$.
        The latter configuration is adopted throughout our network.
        These two convolution layers are designed to have zero-padding and 1-stride
        (in what follows ``same convolution''),
        so that their outputs have the same dimension to their inputs.
  \item Then, the feature fields are convolved by 128 $3^3$-filters, but using a stride of 2.
        Therefore, the outputs are reduced to the size of $16^3$.
        In this step we use the convolution (stride=2)
        to effectively decrease the dimensions of the feature maps, and thus
        reduce the number of parameters to learn and
        the amount of computation performed in the network.
  \item To further extract features and compress them,
        the $16^3$-voxel feature fields are then processed by two same convolution and one convolution (stride=2),
        for further feature extraction and compression.
        Here the three layers have as many as 256 filters,
        as we expect more features when entering a deeper-level regime.
  \item The outputs of the previous layers, i.e. 256 $8^3$-voxel fields,
        are passed to two same convolution layers having 512 $3^3$-filters in each,
        to further extracting features.
  \item After that, a series of deconvolution layers are placed to conduct ``inverse convolution''
        and achieve reconstruction.
        The 512 $8^3$-voxel fields are firstly deconvolved by 256 $3^3$-filters to
        produce $16^3$-voxel fields,
        then convolved by 256 $3^3$-filters for further information extraction,
        and finally deconvolved by 128 $3^3$-filters to recover  $32^3$-voxel fields.
        The deconvolution is achieved via transpose convolution layers \footnote{{Transpose convolution layer
              is very similar to the standard convolution layers, but differs in their receptive field; an easy way to
              realize it is to recongize it as the reverse operation of the convolution layers. And
              one can refer to \url{https://keras.io/api/layers/convolution_layers/convolution3d_transpose/} for more details.}} with stride 1.
  \item Finally, the $32^3$-voxel feature fields are passed to six convolution layers without padding
        (in what follows ``valid convolution'').
        In each valid convolution the $3^3$-filters decrease the size of
        the data by 2, so the final output has a shape of $20^3$.
        They are passed to a deconvolution layer with 3 $3^3$-filters and
        stride 1 to build up a $20^3$-voxel cube with three dimensional
        velocity as the final output.
\end{itemize}

In summary, the network is composed of a series of convolution and
deconvolution layers and have a symmetric structure.
It can be generally considered as an encoder network followed by a decoder network.
In this way, it not only identifies features at the pixel level,
but projects the features learned at different stages of the encoder
onto another pixel space.

A lot of detailed designs are adopted to guarantee the performance of the network.
We summarize them as follows:
\begin{itemize}
  \item In the decoder part, we adopted transpose convolution, instead
        of up-sampling, as the deconvolution layer.
        Compared with the latter design, transpose convolution does a much
        better job in dealing with the non-linearities in the fields.
        Based on the same consideration, in the encoder part, we also use
        transpose convolution, instead of max- or mean-pooling, to reduce
        the data.
  \item After each convolution layer we place one BatchNormalization
        (BN) layer and one activation layer.
        The former one is added to prevent the over-fitting of the model,
        reduce the training cost and improve the training speed.
        The latter one, for which we use rectified linear unit (ReLU) $f(x)
          = max(x, 0)$, is crucial for the neural network, since it brings
        non-linearity into the system.
  \item Each deconvolution is followed by a cropping layer, to match
        the shape of the preceding encoder convolutional density field so
        as to meet the concatenate condition.
        We crop the both side with the same pixel to guarantee either side
        has the same weight to the velocity field.
  \item After every deconvolution, we {\it concatenate} the higher resolution feature fields
        from the encoder network with the deconvolved features,
        in order to better learn representations in the following convolutions.
        Since the decoder is a sparse operation, We need to fill in more details from earlier stages.
  \item During the training,
        we randomly shuffled the input training samples of each epoch to
        prevent the effect of overfitting  due to the similarity of adjacent
        fields.
        %
\end{itemize}

\begin{figure*}[htbp]
  \centering
  \includegraphics[width=16cm,height=5cm]{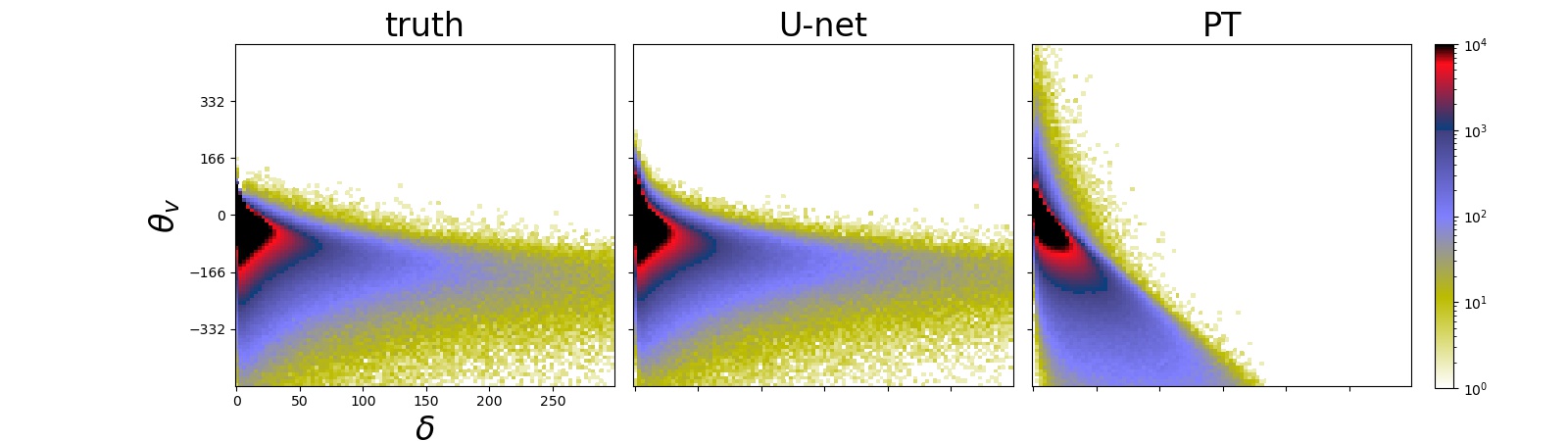}
  \includegraphics[width=16cm,height=5cm]{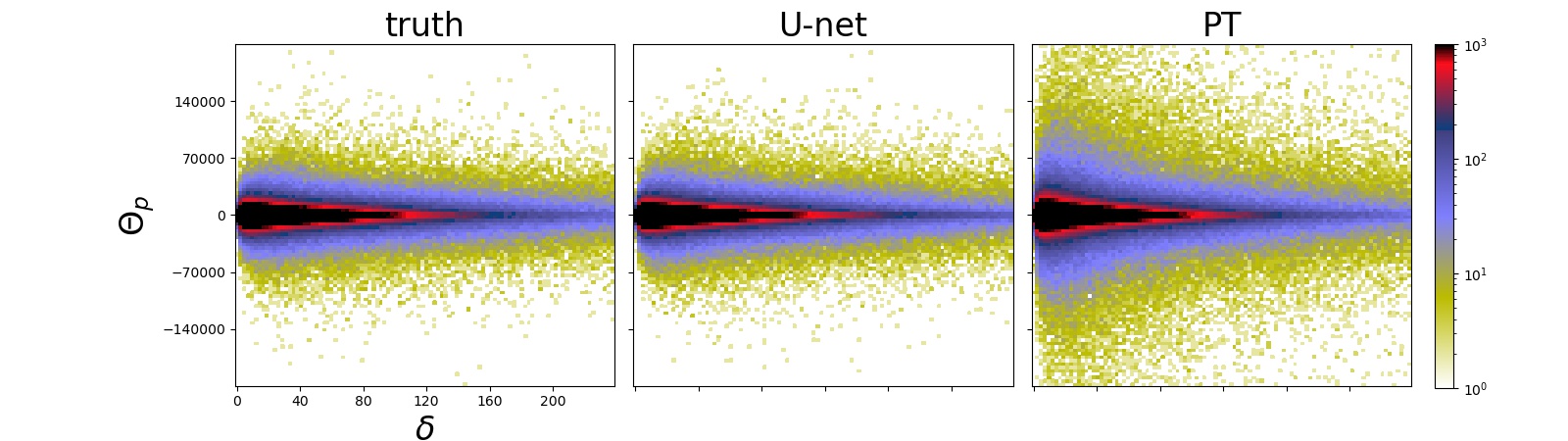}
  \caption{
    Distribution of the velocity divergence $\theta_v$ and the x-direction momentum curl $\Theta_p$,
    along with the density contrast $\delta$.
    From left to right, we show the results calculated using the truth field
    and the fields predicted by the U-net and PT methods,
    in a 480 $h^{-1}$Mpc box with cell-size 2 $h^{-1}$Mpc.
  }
  \label{fig:scattering}
\end{figure*}

\begin{figure*}[htbp]
  \centering

  \includegraphics[width=8cm,height=9cm]{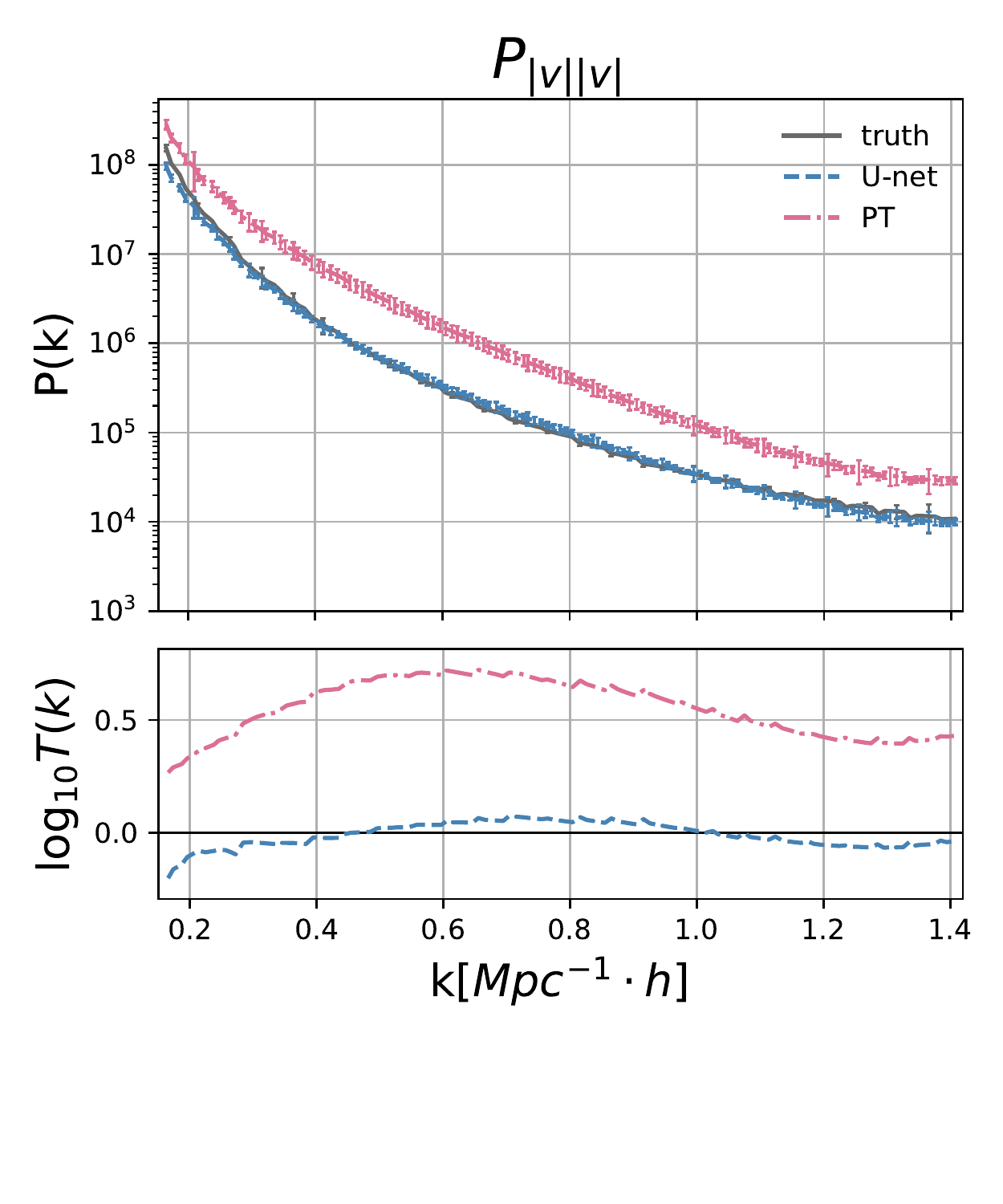}
  \includegraphics[width=8cm,height=9cm]{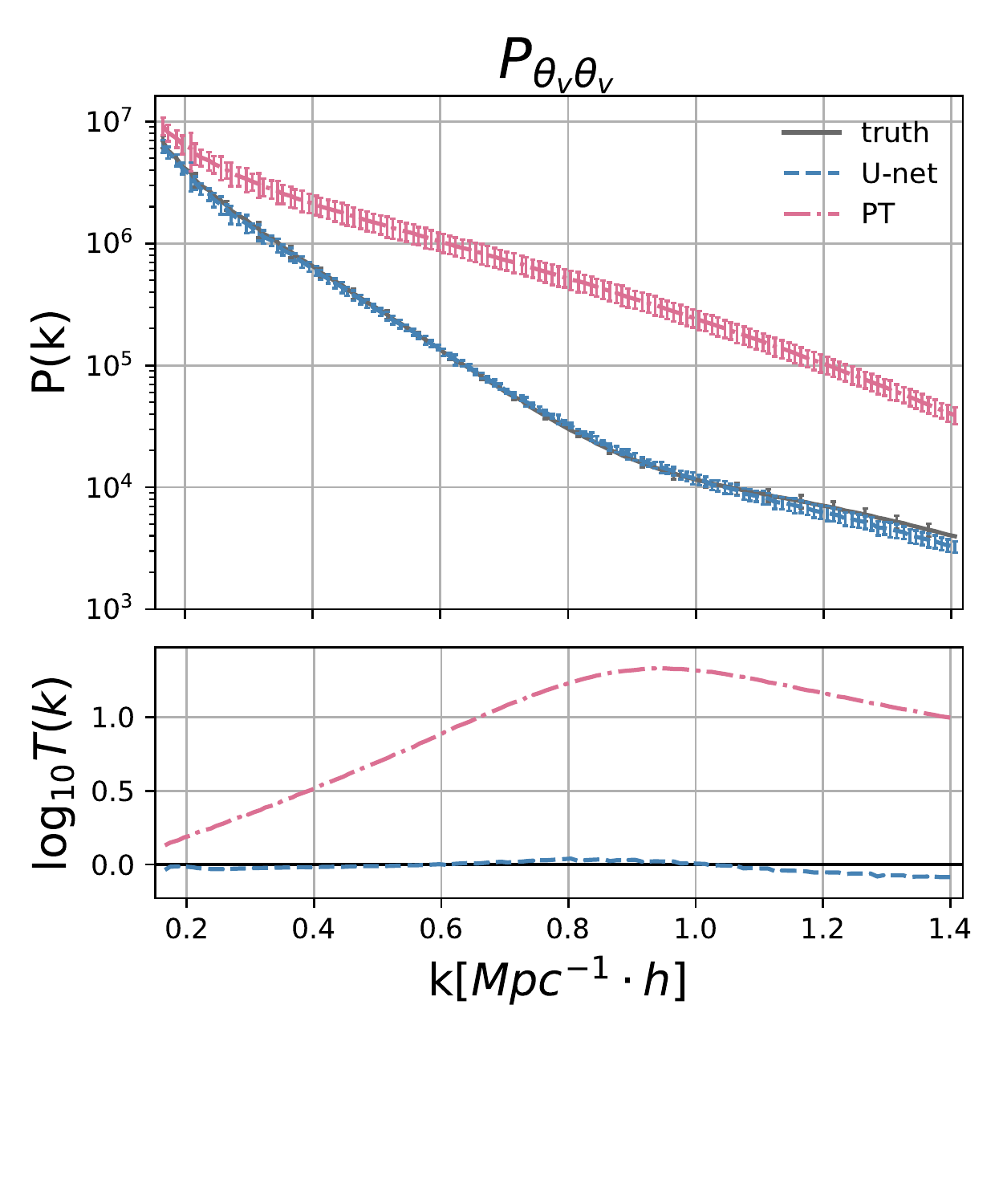}
  \includegraphics[width=8cm,height=9cm]{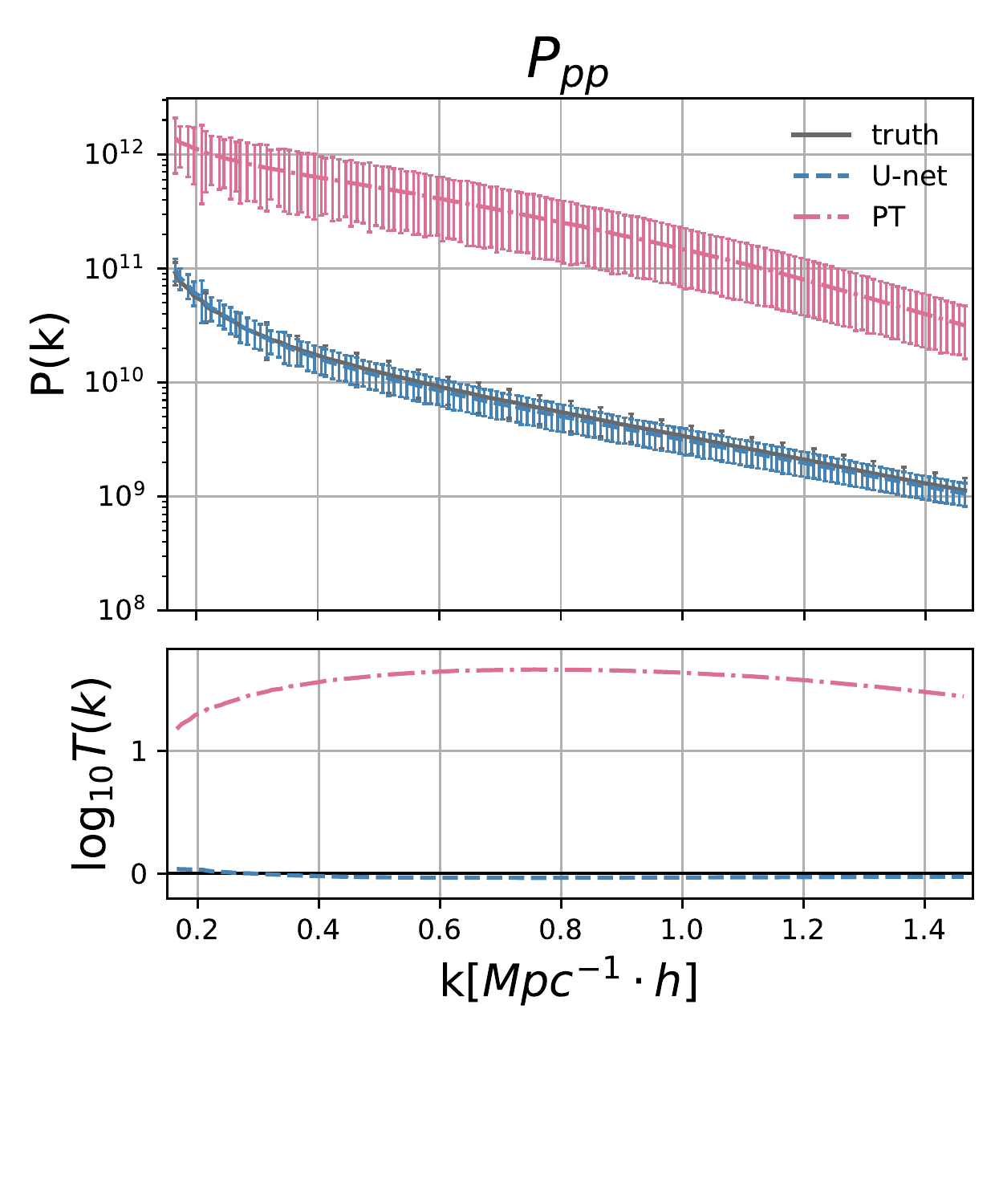}
  \includegraphics[width=8cm,height=9cm]{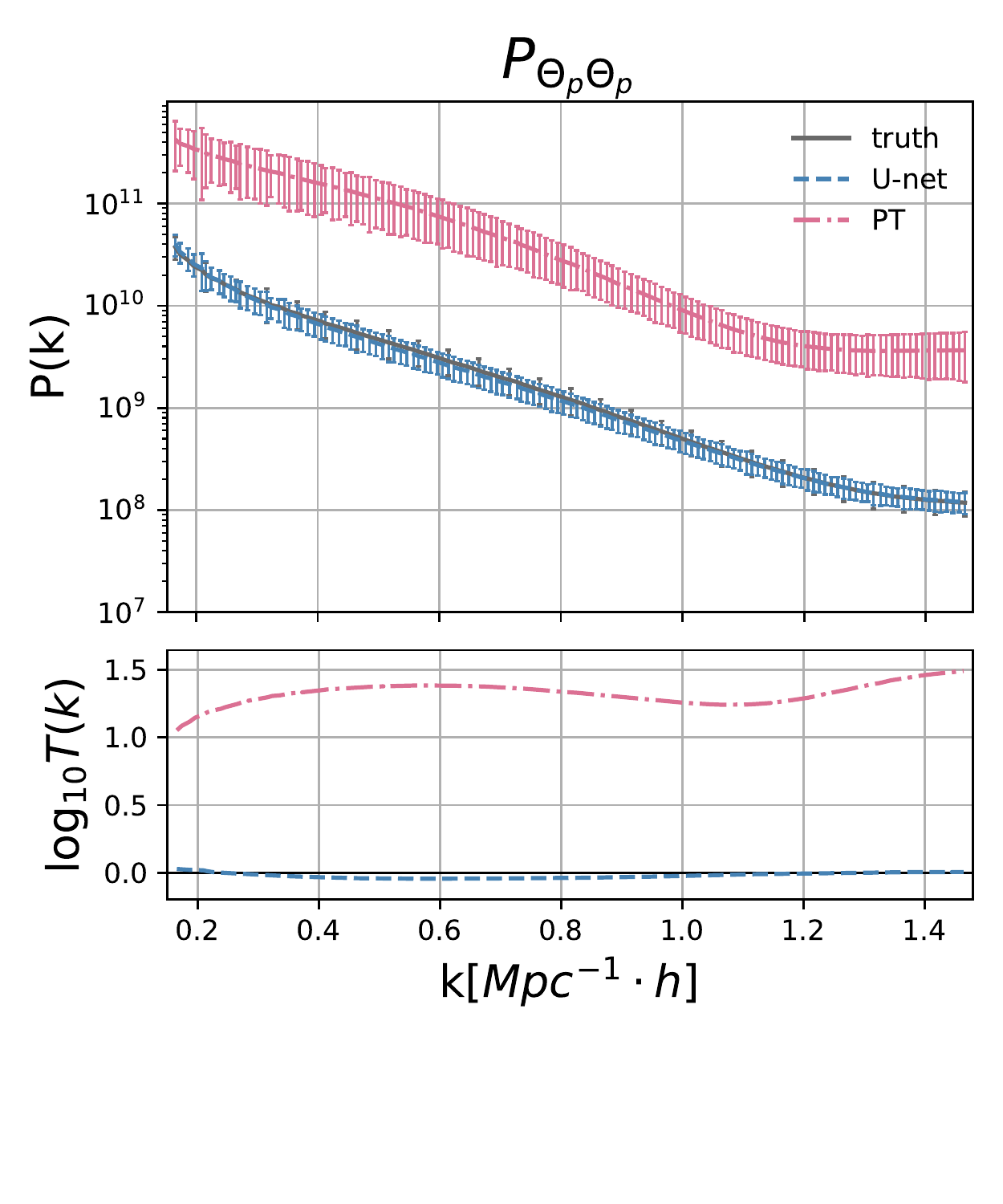}
  \caption{The quantities $P_{|v||v|},\ P_{\theta_v\theta_v},\ P_{{ p}{ p}},\ P_{\Theta_p\Theta_p}$ are
  computed in 64 boxes with the size of $120^3 h^{-1}{\rm Mpc}^3$.
  The gray line, blue line, and pink line represent
  the real field, U-net prediction field, and the linear perturbation theory field,
  respectively.The U-net prediction power spectrum performs much
  better than linear perturbation theory.
  The bottom figures show the deviation of the predicted power
  spectrum to real field.The $y$ axis $\log_{10}T$ is defined as
  $\log_{10}P_{\rm predicted}/P_{\rm true}$.
  All the figures within the range of $0.2\lesssim k\lesssim1.5$ show
  the U-net prediction approximates 0 while the linear perturbation
  theory has great deviation.}
  \label{fig:pk}
\end{figure*}
%
\section{Result}\label{result}



In the following we compare the neural network outputs with the
input truth and the linear perturbation theory expectations.
As mentioned in the previous subsection, in order to suppress the
boundary effect in the training,
the output of the neural network is a $20^3$-voxel field,
located in the center of the $32^3$-voxel input field.
Here we have already put together those sub-cubes into a larger field
(Figure \ref{fig:subcube}).


\subsection{Pixel-to-pixel comparison}

Figure \ref{fig:velocities} shows three slices selected from the {\it
    testing} samples.
They all have a size $40h^{-1}$ Mpc$\times 40\ h^{-1}$Mpc and a thickness $2\ h^{-1}$Mpc.
In all figures, we show the original ``truth'' velocity field,
the predictions of the neural network and the linear perturbation theory,
and also their residuals to the original velocity field.
Plotted in the lower-left corners are the density fields
based on which the velocity fields are derived.

In all cases it is clear that the neural network achieves
a better performance than the linear perturbation theory:
\begin{itemize}
  \item The linear perturbation theory works well in the regime
        where the density and velocity is low
        ({\em e.g.}, see the lower-right corner of the middle and lower panels).
        In the lower-right corner of the lowest panel,
        the performance of the perturbation theory is
        even better than the neural network,
        possibly because the latter puts most effort
        on predicting the non-linear regions.
  \item The linear perturbation theory completely fails
        in the non-linear regions with relatively large density and velocity.
        But the neural network still works well in these regions.
  \item The most interesting cases are those corresponding to
        merging situtations where two regions with opposing bulk velocities
        collide into each other.
        This is shown in he lower-left part of the uppermost panel, the
        upper-left corner of the middle panel,  and the left part of the lowest panel.
        While in these regions the perturbation theory completely fails,
        the neural network still works well in reconstructing
        the velocities.
\end{itemize}

To quantify the performance of the neural network,
for all slices we plot the corresponding histograms of
$|v|$, $|v_{\rm redisual}|/|v_{\rm true}|$,
and $\cos \theta$, where $\theta$ is the angle between
the original and the predicted velocities.

We find the neural network correctly recovers the distribution of $|v|$.
However the linear perturbation theory tends to over-predict the velocity
in the dense regions. In the three slices,
when checking the distribution of $|v|$,
the original fields give
\begin{equation}
  |v|=428\pm188,\ 186\pm103,\ 432\pm208\ \rm km/s\ (\rm original),
\end{equation}
while the neural network predictions give
\begin{equation}
  |v|=376\pm194,\ 177\pm106,\ 368\pm193\ \rm km/s\ (\rm U-net),
\end{equation}
In comparison, the linear perturbation theory predictions are
\begin{equation}
  |v|=649\pm400,\ 386\pm354,\ 797\pm576\ \rm km/s\ (\rm PT).
\end{equation}

Comparing $|v_{\rm residual}|$, the neural network results are
\begin{equation}
  |v_{\rm residual}| =126\pm72,\ 65\pm41,\ 152\pm70\ \rm km/s\ (U-net)
\end{equation}
while the linear perturbation theory yields
\begin{equation}
  |v_{\rm residual}| =281\pm349,\ 298\pm299,\ 407\pm488\ \rm km/s\ (PT)
\end{equation}
The latter results are much worse.
The residual velocities of the neural network results are $3-4$
times smaller than the linear perturbation theory results.

Finally, the neural network perfoms better than linear theory
in predicting the directions of the flows.
They have $|\cos \theta| = 0.93\pm0.19,\ 0.9\pm0.27$, $0.89\pm0.26$
for the three slices, while in the case of linear perturbation theory the results are
$0.92\pm0.26,\ 0.49\pm0.69$, $0.92\pm0.22$.
The neural network results are closer to $1$ and with a smaller
standard deviation.

Similar result can be seen in Fig.\ref{fig:momentum}.
In the non-linear regime, linear perturbation theory completely fails,
while the U-net architecture can still correctly recover the momentum.
When checking the distribution of $|p|$,
the original fields give
\begin{equation}
  |p|=2371\pm4613\ \rm km/s\ (\rm original),
\end{equation}
while the neural network predictions give
\begin{equation}
  |p|=2415\pm4361\ \rm km/s\ (\rm U-net),
\end{equation}
In comparison, the linear perturbation theory predictions are
\begin{equation}
  |p|=1030\pm2318\ \rm km/s\ (\rm PT).
\end{equation}
In the middle panel, we show that the neural network also performs much better
in reconstrucing the curl of the momentum field.

Another important quantity to characterize is the divergence of the
velocity field, given its relevance to study superclusters and the
cosmic web \cite{2012MNRAS.425.2049H,2020MNRAS.tmpL.221P}.
So we analyze the divergence of the velocity field predicted by the
neural network and compare it with the linear perturbation theory.
We find again that the neural network outperforms linear perturbation
theory.
In Fig.\ref{fig:momentum}, the divergence of velocity field
predicted by the neural network is similar to the real one, while the
linear perturbation theory has a larger variance.

In addition, we also made a cell-to-cell comparsion of
the $\delta$-$\theta_v$ and $\delta$-$\Theta_p$ distribution
of the truth field and the U-net or PT predicted fields,
in a 480 $h ^ {- 1} $Mpc box with cell-size 2$h^{-1}$ Mpc.
Figure \ref{fig:scattering} shows that, the scattering pattern of the U-net predicted field is basically consistent with that of the truth field.
In comparison, the PT method leads to a significantly wrong $\delta$-$\theta_v$ distribution,
and also seriously overpredicts the curl value of many cells.

\subsection{Power Spectrum}

We now proceed to check the clustering properties of the fields.
The most commonly used statistics  in cosmological studies
are the two-point correlation function measured in configuration space,
or the power spectrum measured in Fourier space.
In what follows, we compute the two-point correlation function and power spectra of specific quantities
defined as
\begin{equation}
  \begin{aligned}
    \xi_{AA}(|\boldsymbol{r}|) & =\left\langle\delta_{A}\left(\boldsymbol{r}^{\prime}\right) \delta_{A}\left(\boldsymbol{r}^{\prime}+\boldsymbol{r}\right)\right\rangle \\
    P_{AA}(|\boldsymbol{k}|)   & =\int \mathrm{d}^{3} \boldsymbol{r} \boldsymbol{\xi}(r) e^{i \boldsymbol{k} \cdot \boldsymbol{r}}
  \end{aligned}
\end{equation},
where the angle bracket represents the average of the whole sample,
and $A$ denotes the physical quantities we choose to investigate.
In this analysis, the following power spaectrum are taken into account,
\begin{equation}
  P_{|v||v|},\ P_{\theta_v\theta_v},\ P_{{ p}{ p}},\ P_{\Theta_p\Theta_p}
\end{equation}
where $| v | = \sqrt{v_ {x} ^{2} + v_ {y}^{2} + v_ {z}^{2}}$,
$\theta_v \equiv \triangledown\cdot {\bf v}$, ${ p = |v|\delta}$ is the momentum and $\Theta_p \equiv \triangledown \times {\bf p}$ .
In order to compare the difference between the reconstructed field and the actual field, we define
\begin{equation}
  T(k)=\frac{P_{\text {predicted }}(k)}{P_{\text {true}}(k)}
\end{equation}
to characterize the difference between the reconstructed and true fields.
All measures are conducted in $(120 h^{-1} {\rm Mpc})^3$-boxes constructed from the testing samples.

Figure \ref{fig:pk} shows the $ P_{|v||v|},\ P_{\theta_v\theta_v},\ P_{{ p}{ p}},\ P_{\Theta_p\Theta_p}$ of U-net and PT methods and their residuals to the actual power spectrum.
Table \ref{tab:Tk1}\ref{tab:Tk2}\ref{tab:Tk3}\ref{tab:Tk4} compares the ratio of the power spectrum of PT and U-Net in different physical quantities mentioned above to the real power spectrum at $k=0.2, 0.6$ and $1.0$.
When checking the results of $P_{|v||v|}$,
the neural network much better recover its value in the quasi non-linear regime of $k\gtrsim0.2\ h{\rm Mpc}^{-1} $.
In particular,
we find $\lesssim20\%$ discrepancy in $P_{|v||v|}$ within the range of $0.2\ h{\rm Mpc}^{-1}\lesssim k \lesssim 1.4\ h{\rm Mpc}^{-1}$.
The largest discrepancy occurs at $k\simeq 0.272\ h{\rm Mpc}^{-1}$, corresponding to a $T(k)$ of 0.801.
In contrast, the perturbation theory result always has a discrepancy of $T(k)\simeq1.8-5.3$.


Similar results are found when comparing the other two power spectra.
In  $P_{pp}$ we find a $\lesssim8.2\%$ discrepancy within the range of $0.2\ h{\rm Mpc}^{-1}\lesssim k \lesssim 1.4\ h{\rm Mpc}^{-1}$.
The largest discrepancy, at $k\simeq 0.816\ h{\rm Mpc}^{-1}$, corresponds to a $T(k)$ of 0.918, while the perturbation theory result always has a discrepancy of $T(k)\simeq15-47$.

We find $\lesssim18\%$ discrepancy in $P_{\theta_{v}\theta_{v}} $ within the range of $0.2\ h{\rm Mpc}^{-1}\lesssim k \lesssim1.4\ h{\rm Mpc}^{-1}$.
The largest discrepancy, at $k\simeq 1.335\ h{\rm Mpc}^{-1}$, corresponds to a $T(k)$ of 0.818, while perturbation theory consistently exhibits a discrepancy of $T(k)\simeq1.3-22$.

There is a  discrepancy of $\lesssim9.2\%$ in $P_{\Theta_{p}\Theta_{p}} $ within the range of $0.2\ h{\rm Mpc}^{-1}\lesssim k \lesssim1.4\ h{\rm Mpc}^{-1}$.
The largest discrepancy is seen at $k\simeq 0.604\ h{\rm Mpc}^{-1}$ with a $T(k)$ value of 0.907, while perturbation theory has a discrepancy of $T(k)\simeq11-31$.

It is worth noting here that since the outputs of our U-net only have a size of $40 h^{-1}$Mpc,
the sampling spatial sampling limits our ability to accurately recover the large-scale power spectra at $k<0.2\ h{\rm Mpc}^{-1}$.
This can be improved by making corrections on large scales, or simply increase the sizes of the input or output fields.
Since the major focus of this work is to check the capability of the neural network in predicting small-scale, non-linear velocity fields,
we will not discuss this issue in details.

\begin{table}
  \caption{\label{tab:Tk1}  Values of $P_{|v||v|}/P_{\rm true}$, sampled at $k=0.2,\ 0.6 $ and $1.0$}\centering
  \begin{tabular}{c|ccc}
    \hline
    k ($\ h{\rm Mpc}^{-1}$)                               & 0.2   & 0.6   & 1.0   \\
    \hline
    $P_{|v||v|}/P_{\rm true}$, linear perturbation theory & 2.259 & 5.245 & 3.516 \\
    $P_{|v||v|}/P_{\rm true}$, U-net                      & 0.818 & 1.123 & 1.011 \\
    \hline
  \end{tabular}
\end{table}

\begin{table}
  \caption{\label{tab:Tk2}  Values of $P_{pp}/P_{\rm true}$, sampled at $k=0.2,\ 0.6 $ and $1.0$}\centering
  \begin{tabular}{c|ccc}
    \hline
    k ($\ h{\rm Mpc}^{-1}$)                           & 0.2    & 0.6    & 1.0    \\
    \hline
    $P_{pp}/P_{\rm true}$, linear perturbation theory & 20.860 & 44.746 & 43.479 \\
    $P_{pp}/P_{\rm true}$, U-net                      & 1.068  & 0.920  & 0.929  \\
    \hline
  \end{tabular}
\end{table}

\begin{table}
  \caption{\label{tab:Tk3}  Values of $P_{\theta_{v}\theta_{v}}/P_{\rm true}$, sampled at $k=0.2,\ 0.6 $ and $1.0$}\centering
  \begin{tabular}{c|ccc}
    \hline
    k ($\ h{\rm Mpc}^{-1}$)                                             & 0.2   & 0.6   & 1.0    \\
    \hline
    $P_{\theta_{v}\theta_{v}}/P_{\rm true}$, linear perturbation theory & 1.584 & 7.917 & 20.731 \\
    $P_{\theta_{v}\theta_{v}}/P_{\rm true}$, U-net                      & 0.956 & 0.996 & 1.011  \\
    \hline
  \end{tabular}
\end{table}

\begin{table}
  \caption{\label{tab:Tk4}  Values of $P_{\Theta_{p}\Theta_{p}}/P_{\rm true}$, sampled at $k=0.2,\ 0.6 $ and $1.0$}\centering
  \begin{tabular}{c|ccc}
    \hline
    k ($\ h{\rm Mpc}^{-1}$)                                             & 0.2    & 0.6    & 1.0    \\
    \hline
    $P_{\Theta_{p}\Theta_{p}}/P_{\rm true}$, linear perturbation theory & 14.806 & 24.086 & 18.065 \\
    $P_{\Theta_{p}\Theta_{p}}/P_{\rm true}$, U-net                      & 1.048  & 0.907  & 0.95   \\
    \hline
  \end{tabular}
\end{table}

\section{Discussion and Conclusions}\label{conclu}

In this paper, we applied a deep learning technique to reconstruct the
velocity field from the dark matter density field, which has a
resolution of $2$ $h^{-1}$Mpc.
To this end we implement a ``U-net'' neural network, consisting of 15
convolution layers and 2 deconvolution layers with 48,690,307
parameters. The network maps the $32^3$-voxel input density field  to
velocity and momentum fields having size of $20^3$, so as to avoid boundary effects.

We find that the neural network manages to reconstruct the velocity
and momentum fields and even outperforms the results from linear
perturbation theory.
The superiority of the neural network is more pronounced in regions
where the density is relatively large and the non-linear processes
dominate.  In particular, in regions where mergers take place, linear perturbation theory completely fails,
while the neural network succesfully recovers the velocity structure.

By conducting pixel-to-pixel comparison between the predicted velocity fields and underlying true fields,
we find that the neural network can reasonably recover the
distribution of $|v|$, having discrepancy of $|v_{\rm
  residual}|\lesssim150\ $ km/s, while for the perturbation theory
results we find $|v_{\rm residual}|\simeq 300-400$ km/s. The neural
network also predicts well the directions of the velocities compared  to the true velocities.

When analyzing the clustering properties of the fields, the neural
network can well recover the amplitude and shape of $P_{|v||v|}$,
whose error ranges from 1\% to $\lesssim$10\% within the range of $0.2\lesssim
  k\lesssim1.5$. Similarly, the error of $P_{pp}$ is $\lesssim8.2\%$,
$P_{\theta_{v}\theta_{v}} $ is $\lesssim17\%$ and
$P_{\Theta_{p}\Theta_{p}} $ is $\lesssim9.2\%$ at the range of
$0.2\lesssim k\lesssim1.4$ All these results are much better than the linear perturbation theory results.

As a proof-of-concept study, our analysis demonstrates the ability of
deep neural networks to reconstruct the nonlinear velocity and
momentum fields from density fields.
The neural network can even handle regions of shell-crossing,
which is notoriously difficult within perturbation theory approaches.
At the same time,
there is still much room for improvement in the accuracy of the neural network,
via further optimizing the architecture, enlarging the number of the training samples,
or adding follow-up neural networks to fit the residuals and do corrections.

The reconstructed peculiar velocity fields can be used for a number of studies,
such as BAO reconstructions, RSD analyses, kinematic Sunyaev-Zeldovich
(kSZ), supercluster analysis  and the cosmic web construction.
We will continue to work on this direction so that the machine learning technique
can be reliably applied to real observational data and help us uncover
more of the mysteries of the universe.

\begin{acknowledgments}
  We thank Kwan-Chuen Chan, Yin Li, Jie Wang, Le Zhang and Yi Zheng for helpful discussions.
  This work is supported by National SKA Program of China No. 2020SKA0110401.
  XDL acknowledges the support from the NSFC grant (No. 11803094) and the Science and Technology Program of Guangzhou, China (No. 202002030360).
  CGS acknowledges financial support from the National Research Foundation of Korea (NRF; \#2020R1I1A1A01073494).
  J.E. F-R acknowledges support from COLCIENCIAS Contract No. 287-2016,
  Project 1204-712-50459.
  Y.W. is supported by NSFC grant No.11803095
  and NSFC grant No.11733010.
  We acknowledge the use of Tianhe-2 supercomputer.
  We also acknowledge the use of the
  Kunlun cluster, a supercomputer owned by the School of
  Physics and Astronomy, Sun Yat-Sen University.
\end{acknowledgments}

\bibliography{references}

\end{document}